\documentstyle[12pt]{article} 
\begin{document}

\begin{center}
{\Large \em EXPLOSIVE COSMOGONY AND \\}
{\Large \em  THE 
  QUASI-STEADY STATE COSMOLOGY\footnote{Invited paper presented at IAU
Symposium No. 183 ``Cosmological Parameters and evolution of the Universe''
read August 21, 1997 in Kyoto, Japan (Kluwers Academic Publishers, 
in press)}}       
\end{center}

\begin{center}
{\bf G. BURBIDGE \\} 
{\bf University of California, San Diego\\
\bf Center for Astrophysics \& Space Sciences \& Dept. of Physics\\
\bf La Jolla, CA 92093-0424}         
\end{center}

\begin{abstract}
Arguments and evidence against the hot big bang model are summarized.  The
observations which point to an explosive cosmogony and the quasi-steady 
state cosmology are outlined.
\end{abstract}

\section{The Historical Setting}

Modern cosmology began with the realization that there were solutions to
Einstein's theory of gravity discovered by Friedmann and Lemaitre which
when combined with the redshift distance relation of Hubble and others
could be interpreted as showing that we live in an expanding universe.
By 1930, the scientific establishment and many of the lay public believed
this.  It was then only elementary logic to argue that if time reversal
was applied, the universe must originally have been so compact that we
could talk of a beginning.  Lemaitre tried to describe this state as the
``Primeval Atom.''  For a decade or so after the war, Gamow, Alpher and
Herman and other leading physicists explored this dense configuration
trying to make the chemical elements from protons and neutrons.  They
soon learned that this was not possible because of the absence of stable 
masses of five and eight, but they also realized that \underline{if} 
such an early
stage had occurred the universe would contain an expanding cloud of radiation
which would preserve its black body form.  Dicke and his colleagues in
Princeton rediscovered this idea and decided to try and detect the 
radiation.
Penzias and Wilson found such a radiation field, and COBE has 
demonstrated that
it has a perfect black body form out to radio wavelengths.  This history
of the discovery together with the fact that the light elements D, 
He$^{3}$ and He$^{4}$ in about the right amounts can be made in a hot big
bang has led to the widely held, but simplistic view, that the standard
cosmology -- the hot big bang -- is correct.

As the belief in this theory has grown, with some of its popularity coming
from the fact that a beginning is a main theme of western religion, it
has become progressively harder to argue against it.

\section{Real Objections to the Big Bang Model}

\subsection{THE EARLY DISCOVERY AND THE He/H PROBLEM}  

What was not properly understood by those who first discussed the 
early universe was that McKellar had already discovered the microwave
radiation in 1941 (he obtained a temperature of 2.3$^{\circ}$K setting
as a lower limit 1.8$^{\circ}$, and as an upper limit 3.4$^{\circ}$K for
the interstellar temperature) (McKellar 1941). 
 It is also clear that the only reason why
the physicists decided to invoke a dense configuration in an early universe 
was to find a place with a plentiful source of neutrons.  Since the model
failed to explain the building of nearly all of the 
chemical elements (which we now know
following Hoyle were made in the stars) the model might well have been 
dropped.  This is especially clear when it is also pointed out that in 
the 1950s both Bondi, Gold and Hoyle (1955) and independently,  
Burbidge (1958) pointed out 
that if the observed abundance of He was obtained by 
hydrogen burning in stars, there must have been a phase in the history
of the universe when the radiation density was much higher than the energy
density of starlight today.  The very striking fact is that if we 
suppose that if $\rho$ is the density of visible matter in the universe,
with a value of about $3 \times 10^{-31} gm \; cm^{-3}$, and that the 
He/H ratio by mass in it is 0.244, then the energy which must have been
released in producing He is $4.39 \times 10^{-13} erg \; cm^{-3}$.  
If this energy is thermalized, the black body temperature 
iturns out to be $T = 2.76^{\circ}$K.
This value is astonishingly close to the value of 2.73$^{\circ}$K 
observed by COBE.

This simple agreement of two measured quantities makes no allowance for the 
expansion of the universe that must necessarily have taken place during
the production of helium, which would act to reduce the temperature.
However, it does show that unless it is dismissed as
a coincidence which all big bang believers must do, there is likely to be an
explanation of the microwave radiation in terms of straight forward
astrophysics involving hydrogen burning in stars.

It can be argued therefore that this line of reasoning completely
refutes the popular view that the discovery of the microwave radiation
is proof that a big bang occurred.

The usual rebuttal to this argument is that it is the blackbody nature
of the radiation that is important, not the value of the temperature,
and that in any alternative scheme it is the thermalization process of
the radiation that is the weak link.

The counter to this argument is that in the standard model 
generation of the black body 
radiation is traced to the decay of the false vacuum energy in the
inflationary scenario.  But as we shall show this whole approach is a 
gross extrapolation beyond known physics.

\subsection{THE ARBITRARY NATURE OF THE PHYSICS IN THE EARLY UNIVERSE}

This can be seen by 
reversing the time axis associated with the expansion.  
As the universe shrinks the radiation energy
begins to dominate the matter and ultimately  the matter is 
broken down into quarks. 
We now move out of the realm of known physics.  A
further contraction by a factor of about $10^{10}$ is invoked leading
to what is called a ``phase transition'' in which everything is converted 
into a new kind of so-called scalar particle.  These scalar particles are 
supposed to interact together to produce what is described as a ``false
vacuum'' maintaining positive energy at all costs.  This false vacuum
consumes space-time in a process of deflation -- this is the inflation
epoch of Guth and Linde when time is reversed.  The consuming of spacetime
leads to what?  To a quantum transition to somewhere else!

If these words are not understandable to the layman it is because they are
not understandable to physicists either, however ``nice'' the idea of 
inflation may be.  And if not inflation what else?  The problem is the 
resistance to the idea (of Dirac) of particles of negative energy.  While
the energetics are still outside the realm of known physics, the existence
of a negative energy field will permit entirely new positives to form with the new positives compensating those of negative energy with what we can
refer to as creation events in which energy is conserved.  This approach is
already preferable.

The great importance of it is that there is much observational evidence
pointing to such creation events.  This evidence 
is the cosmogonical basis for
the cosmological quasi-steady state theory which Dr. Narlikar will
outline.

In the classical big bang cosmology all of the discrete objects are supposed
to have arisen from density fluctuations in the early universe.  While
there is no direct evidence at all for this, evidence of the other kind is 
widespread.
 None of this evidence is understood within the framework of the big
bang cosmology. 

\section{The New Observational Evidence}

Starting about forty years ago, Ambartsumian (1958, 1965) proposed that 
groups and clusters of galaxies that appear to be expanding are doing 
just that.
They are systems of positive total energy.  This idea was resisted by Oort
and others on the ground that the galaxies must be very old, but if they were
being ejected in expanding associations this could not be correct.  
To bind the
groups and clusters dark matter was invoked.  
This was one of the early arguments for the
presence of dark matter.

For clusters which are 
really relaxed and obviously stable this is a perfectly good argument, 
but for many, if not the majority of clusters 
 which show every sign of instability it
seems likely that the virial does not hold and they are coming apart.  There
is only limited 
evidence, of course, that galaxies in general are $\sim 10^{10}$ years
old.  We now accept the fact that many galaxies have very young components,
and perhaps whole galaxies may be young.

Also in the 1960s, it became clear that the nuclei of galaxies often give
rise to explosive outbursts (Burbidge, Burbidge \& Sandage 1963).  
The most powerful of
these are the radio outbursts which can generate $10^{60} - 10^{62}$ ergs
largely in the form of relativisitic particles.  With the discovery of the
quasi-stellar objects came the evidence that many of the radio emitting
QSOs lie so close to comparatively nearby galaxies that the configurations
cannot be accidental, although the QSOs have much larger redshifts then
the galaxies.  Thus it follows that the QSOs must have been ejected from
the galaxies and must have large intrinsic redshift components.  Evidence of 
this kind has been obtained in profusion over the last twenty-five years
by Arp, ourselves and others (cf Burbidge, Burbidge \& Sandage 1963, 
Arp 1987, Burbidge 1996).

Most recently clear cut evidence showing that X-ray emitting QSOs are being
ejected from nearby active galaxies (NGC 4258, 1068, 3516, 5548, etc.)
has been obtained.  Arp will describe some of the new results.

How have these observations been explained in the classical picture.  Two
lines of attack hae been developed.  The paradigm put forward to explain
the activity in galactic nuclei is that the energy is gravitational in
origin and is generated from matter falling into the center which
contains a massive black hole and an accretion disk (cf Rees 1984).  This
paradigm is never tested but continuously asserted.  In our early 
calculations (Hoyle et al. 1964) we showed how implausible this was.
As far as the evidence for QSOs etc. ejected from galaxies is concerned
the position that is taken is that the data are all suspect and therefore
need not be explained.

On the other hand, we accept the 
observational evidence of expanding groups and
clusters, and ejection phenomena involving high energy events from
the centers of radio galaxies and many other active galaxies which shows that
dense objects are ejected from the nuclei of already existing galaxies.

Thus we believe that galaxies beget galaxies, not that they are made from
initial density fluctuations in a big bang universe.

The cosmological model which is 
based on this cosmogony is 
the quasi-steady state cosmology.  This is the model developed by Hoyle,
Burbidge and Narlikar which has been published in a series of papers
recently (Hoyle, Burbidge and Narlikar 1993, 1994ab, 1995):

In this theory there can be particles with negative energy.  The particles
exert a negative pressure.  This negative pressure produces the expansion
of the universe, not an initially assumed explosion.  The universe goes
through repeated cycles of expansion and contraction, each expansion
driven by negative pressure inside many localized regions (galaxies) 
where a large amount of creation occurs.

These cycles with a total period of about $40 \times 10^{9}$ years
are superposed on an overall slow expansion with a time scale $\sim 10^{12}$
 years.

Dr. Narlikar will discuss the ways in which we can understand the
abundances of the light elements and the microwave background radiation
within the framework of this theory. 




\section{References}
{\small


\begin{thebibliography}{}  
\bibitem{} Ambartsumian, V. A. 1958, Solvay Conf. Report,(ed R. Stoops)
Bruxelles. 
\bibitem{} Ambartsumian, V.A. 1965,``Structure and Evolution of Galaxies''
Proc. 13th Solvay Conf. on Physics, University of Brussels 
(Wiley Interscience)
\bibitem{} Arp, H.C. 1987, ``Quasars, Redshifts \& Controversies'' 
(Interstellar Media, Berkeley). 
\bibitem{} Bondi, H., Gold, T. \& Hoyle, F. 1955, Observatory Mag., 75, 80.
\bibitem{} Burbidge, G. 1958, P.A.S.P., 70, 83. 
\bibitem{} Burbidge, G. 1996, A\&A, 309,9. 
\bibitem{} Burbidge, G., Burbidge, M. \& Sandage, A.R. 1963, Rev. Mod. Phys.
35, 947. 
\bibitem{} Hoyle, F., Burbidge, G. \& Narlikar, J.V. 1993, ApJ, 410, 437. 
\bibitem{} Hoyle, F., Burbidge, G. \& Narlikar, J.V. 1994a, MNRAS, 267, 1007. 
\bibitem{} Hoyle, F., Burbidge, G. \& Narlikar, J.V. 1994b, A\&A, 289, 729. 
\bibitem{} Hoyle, F., Burbidge, G. \& Narlikar, J.V. 1995, 
Proc. Roy. Soc., A, 448, 191.
\bibitem{} Hoyle, F., Fowler, W.A., Burbidge, G., \& Burbidge, M. 
1964, ApJ, 139, 909. 
\bibitem{}
McKellar, A. 1941, Pub. Dom. Astrophy. Observatory, Vol. 7, No. 15 
\bibitem{} Rees, M.J. 1984, ARA\&A, 22, 471. 
\end{thebibliography}
\end{document}